\documentclass[a4paper,11pt]{article}
\usepackage{jinstpub} 
\usepackage{siunitx}

\proceeding{Light Detection In Noble Elements - LIDINE 2023\\
September 20-22, 2023\\
Madrid}

\title{Towards a fiber barrel detector for next-generation high-pressure gaseous xenon TPCs}

\collaboration[c]{on behalf of the NEXT collaboration}

\author{S.R. Soleti}
\affiliation{Donostia International Physics Center\\
Manuel Lardizabal Ibilbidea, 4, 20018 Donostia, Spain}

\emailAdd{roberto.soleti@dipc.org}

\abstract{The NEXT (Neutrino Experiment with a Xenon TPC) project is an international collaboration aimed at finding evidence of neutrinoless double beta decay using gaseous xenon. The current phase of the project involves the construction and operation of NEXT-100, which is designed to hold 100 kg of xenon at 15 bar and is expected to start commissioning in the first quarter of 2024. NEXT-HD will be a tonne scale experiment following NEXT-100 and will incorporate a symmetric design, with one cathode and two anodes. For this detector, the collaboration is considering to implement a barrel of wavelength-shifting fibers read-out by silicon photomultipliers to measure the energy of the particles interacting in the gaseous xenon. In this document, we will discuss the characteristics of this approach and provide an update on the related R\&D efforts.}


\begin{document}
\maketitle
\flushbottom

\section{A fiber barrel detector for a high-pressure gaseous xenon TPC}
\label{sec:intro}
The NEXT collaboration has successfully developed the high-pressure gaseous xenon time projection chamber with electroluminescence technology (HPXe-EL) for neutrinoless double beta decay searches ($0\nu\beta\beta$)~\cite{NEXT:2012rto, NEXT:2018rgj}. The current phase of the experiment is called NEXT-100 and the corresponding detector is designed to hold 100 kg of xenon at 15 bar~\cite{NEXT:2012zwy}. Its construction is being completed and will start its commissioning phase in January 2024. 
NEXT-100 adopts the \emph{SOFT} (Separated Optimized FuncTion) design, where the energy of the particles interacting in the gaseous xenon is measured by a plane of photomultipliers (PMTs) placed on the cathode side and the tracking of the ionization electrons is performed by an array of silicon photomultipliers (SiPMs) placed on the anode side. 

The adoption of PMTs to measure the energy presents two main limitations: they cannot withstand high pressures, thus introducing mechanics complication in the experiment design, and they represent a significant radioactive background, which is especially important for $0\nu\beta\beta$ searches~\cite{MartinezLema:2018lix}. 

\begin{figure}[htbp]
\centering
\includegraphics[width=.4\textwidth]{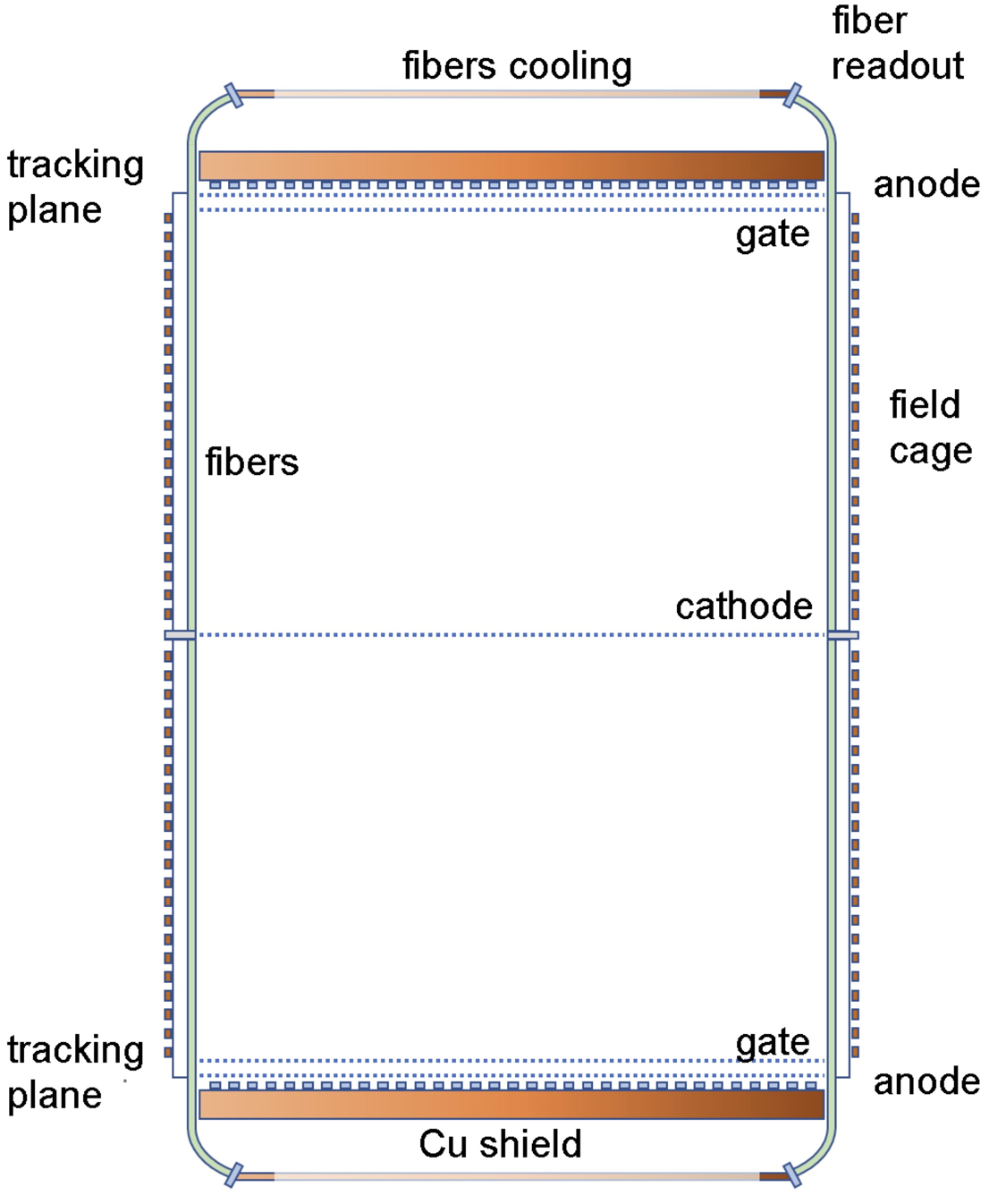}
\qquad
\includegraphics[width=.3\textwidth]{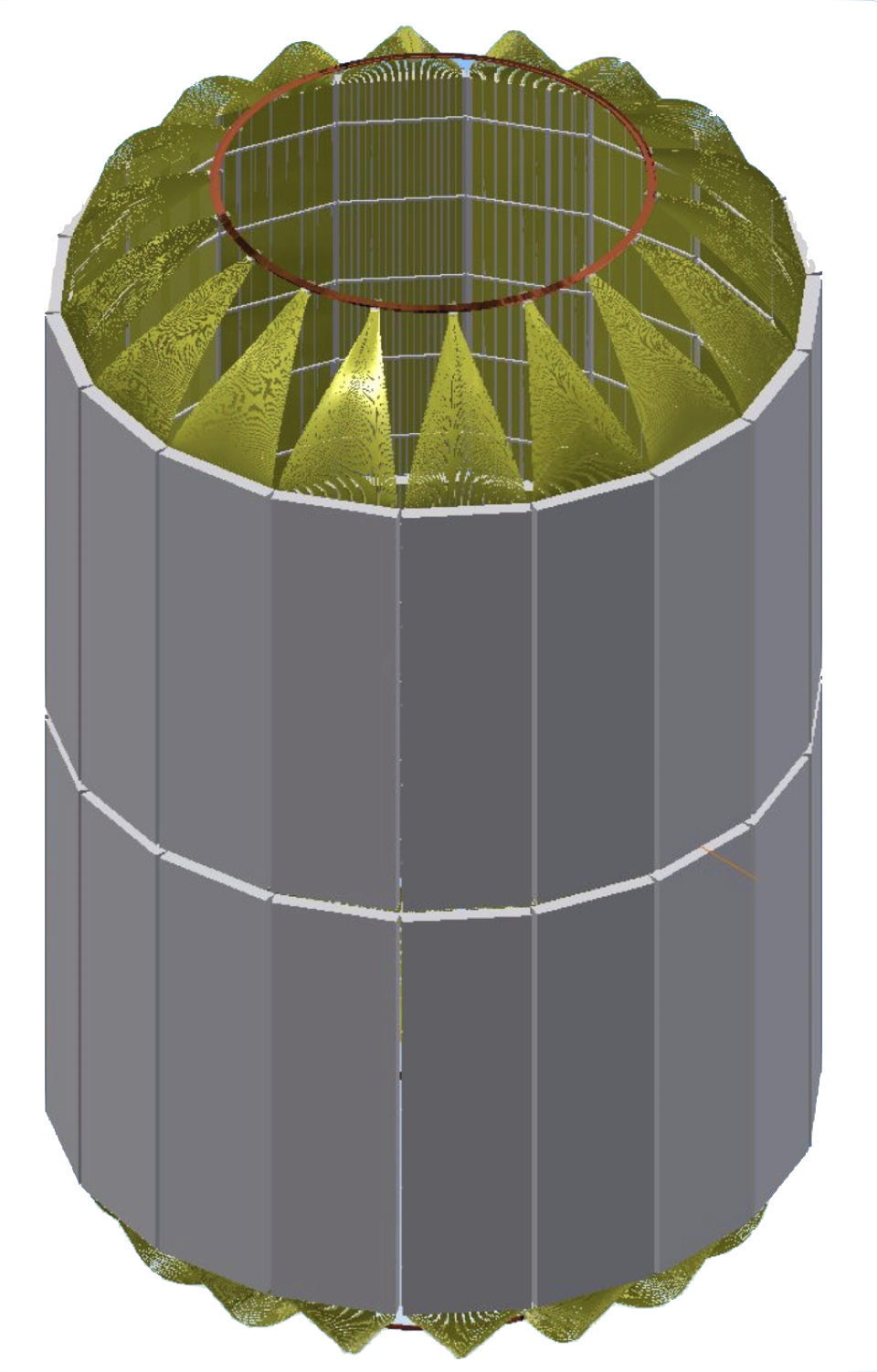}
\caption{\emph{Left:} Schematic of the symmetric design being considered for the NEXT-HD detector. The cathode is placed at the center and two tracking planes are placed on the two anode sides. A barrel of WLS fibers surrounding the detector measures the energy of the event. \emph{Right:} a CAD drawing of the fiber barrel. The fibers are placed on polytetrafluoroethylene (PTFE) panels and the light is detected by SiPMs placed on a copper ring for cooling purposes.\label{fig:detector}} 
\end{figure}

For the next iteration of the experiment, the collaboration plans to build a tonne-scale experiment called NEXT-HD \cite{NEXT:2020amj}. For its design, a \emph{symmetric} layout is being explored, with the cathode placed in the center and two tracking planes placed on the two anode sides. With this configuration, the active region is doubled for the same cathode voltage, and the diffusion will be reduced given the shorter drift distance. The energy is measured by a barrel of wavelength-shifting (WLS) optical fibers covering the side of the detector (see figure~\ref{fig:detector}). The light of the fibers is detected by SiPMs, whose dark noise is suppressed by placing them in contact with a cooled copper ring. 

These proceedings focus on the R\&D effort dedicated to the fiber barrel and are organized as follows: the technology options being considered are discussed in section~\ref{sec:tech}, the design of a prototype of the NEXT-HD detector is detailed in section~\ref{sec:hddemo}, the current experimental R\&D effort is described in section~\ref{sec:rd}, and a summary of the current and future efforts for the project is presented in section~\ref{sec:summary}.

\section{Technology choice\label{sec:tech}}
\subsection{Wavelength-shifting fibers}
A WLS fiber barrel for the NEXT experiment must satisfy three main requirements: (1) be able to detect scintillation photons from $^{83\mathrm{m}}$Kr decays, which release a total energy of 41.5 keV and are used for detector calibration, (2) be radiopure enough to not increase the radioactive budget of the detector, (3) be mechanically viable and work in a high-pressure vessel filled with gaseous xenon. 

\begin{figure}[htbp]
\centering
\includegraphics[width=.98\textwidth]{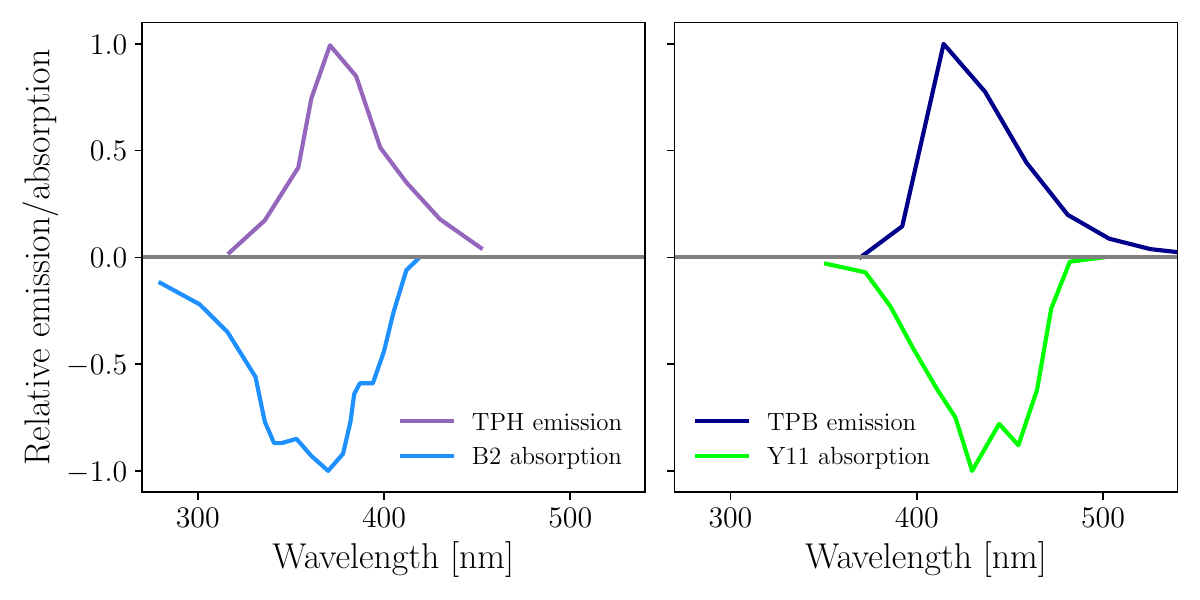}
\caption{Emission spectrum of the wavelength-shifting coating (TPH~\cite{Akimov:2012lbl} and TPB~\cite{Benson:2017vbw}) and absorption spectrum of the corresponding fiber (Kuraray B2 and Kuraray Y11).\label{fig:fibers}} 
\end{figure}

The current baseline design employs round WLS fibers with a diameter of 1 mm coated with a wavelength shifting material, whose goal is to shift the $\sim$170~nm photons emitted by the excited xenon gas to a wavelength large enough to be absorbed by the fiber. The top of the fiber away from the photosensor is coated with an aluminum mirror, which provides a reflectivity of $\sim75\%$~\cite{Saraiva:2004cn}. Two options for the WLS fibers are being considered: blue-to-green fibers (e.g. Kuraray Y11~\cite{kuraray}) coated with tetraphenyl butadiene (TPB) and UV-to-blue fibers (e.g. Kuraray B2) coated with p-terphenyl (TPH).
In both cases, the emission spectra of the WLS materials are well matched to the respective fiber absorption lengths, as shown in figure~\ref{fig:fibers}. TPB has a long history of being employed as a wavelength-shifting material in particle physics experiments. As an example, a barrel of WLS fibers coated with TPB and submerged in liquid argon has been successfully implemented by the GERDA experiment as a Compton veto~\cite{Janicsko-Csathy:2010uif, Csathy:2016wdy}. TPB does not significantly affect the purity of gaseous xenon and it has already been used to coat the light tube and the tracking plane of the NEXT-White and NEXT-100~\ detectors. 

P-terphenyl represents an attractive alternative, given its high quantum efficiency~\cite{Akimov:2010six, Akimov:2012lbl}. Its use as a coating material has been pioneered by the DUNE collaboration for the X-ARAPUCA device~\cite{Brizzolari:2021akq}. However, the literature provides somehow conflicting measurements~\cite{DeVol:1994bi} and a dedicated R\&D effort is necessary to assess its performances as a coating material for optical fibers and its behavior in gaseous xenon. 

\subsection{Photosensors}
In the baseline design of the barrel fiber detector, the light exiting the WLS fibers is detected by SiPMs, whose adoption avoids the complications of the operation of PMTs in high-pressure environments. At parity of readout surface, SiPMs are much more compact than PMTs and they can be made of radiopure material, thus improving the radioactive budget of the experiment.
The model being considered is Hamamatsu S13360-6075PE~\cite{hamamatsu}, which has an area of $6\times6$ mm$^2$, thus being able to read bundles of approximately 30 WLS fibers (although the silicon area could cover in principle 36 fibers, packing more than 30 fibers in a squared bundle is mechanically challenging). 

The photon detection efficiency of this model is well matched to the emission spectra of Y11 and B2 fibers and its emission-weighted average is approximately 45\% in both cases. 

The main limitation of SiPMs for low-backgrounds experiments such as NEXT is represented by their non-negligible dark noise at room temperature. A single S13360-6075PE sensor has a typical dark rate of approximately 2 MHz and the NEXT-HD detector will have approximately 200 photosensors on each side. A preliminary simulation of a barrel fiber detector gives roughly 25 detected photoelectrons for a single $^{83\mathrm{m}}$Kr event, which represents a factor of 2 improvement compared to the NEXT-White energy plane equipped with PMTs. Thus, at a typical sampling rate of 1~\si{\micro\second}, the signal-to-noise ratio is $<0.1$. Dark rate can be reduced by cooling the photosensors (approximately by a factor of 1.5 every 5 degrees \cite{NepomukOtte:2016ktf}). However, the temperature cannot be lower than the condensation point for xenon at 15 bar, which is approximately 230 K. At this temperature, the simulation gives a signal-to-noise ratio $>10$.

\section{The HD-DEMO detector\label{sec:hddemo}}
\begin{figure}[htbp]
\centering
\includegraphics[width=.9\textwidth]{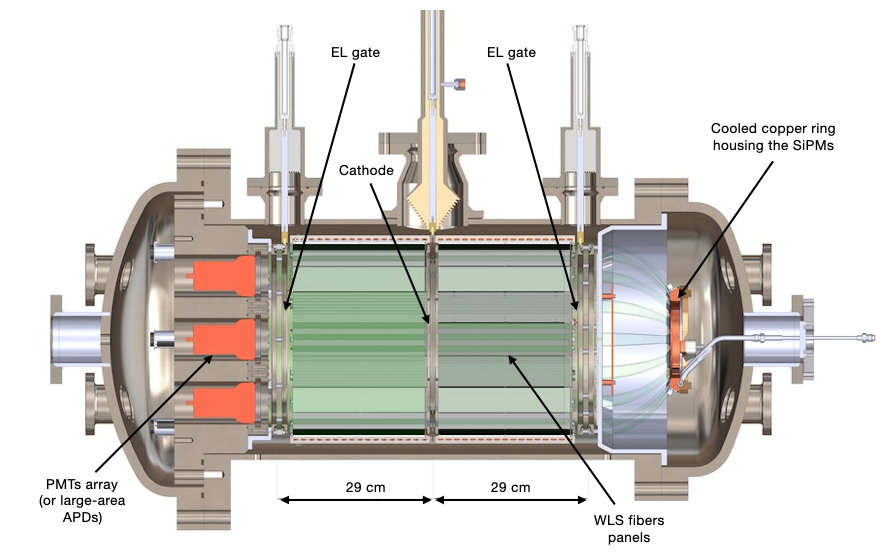}
\caption{The HD-DEMO detector will consist of a symmetric high-pressure gaseous xenon TPC. The energy will be measured by a barrel of WLS fibers read out by cooled SiPMs on one side and by PMTs or large-area APDs on the other side. \label{fig:hd_demo}} 
\end{figure}

The HD-DEMO detector, currently in its final design phase, will be used as prototype to explore the detector technologies of the NEXT-HD experiment. It consists of a symmetric, high-pressure gaseous xenon TPC with a diameter of 35~cm and two drift volumes, each one 29~cm long. A CAD drawing of the detector is shown in figure~\ref{fig:hd_demo}. In one volume, the fibers will be read out by a ring of cooled SiPMs, and in the other one by an array of photomultipliers or large-area APDs, currently being developed. The detector will allow to validate the engineering solutions necessary to cool the SiPMs in a high-pressure gaseous environment and to compare their performances with the ones of alternative photosensors. Its construction is expected to start in 2024.

\section{Fiber barrel R\&D\label{sec:rd}}
\begin{figure}[htbp]
\centering
\includegraphics[width=.52\textwidth]{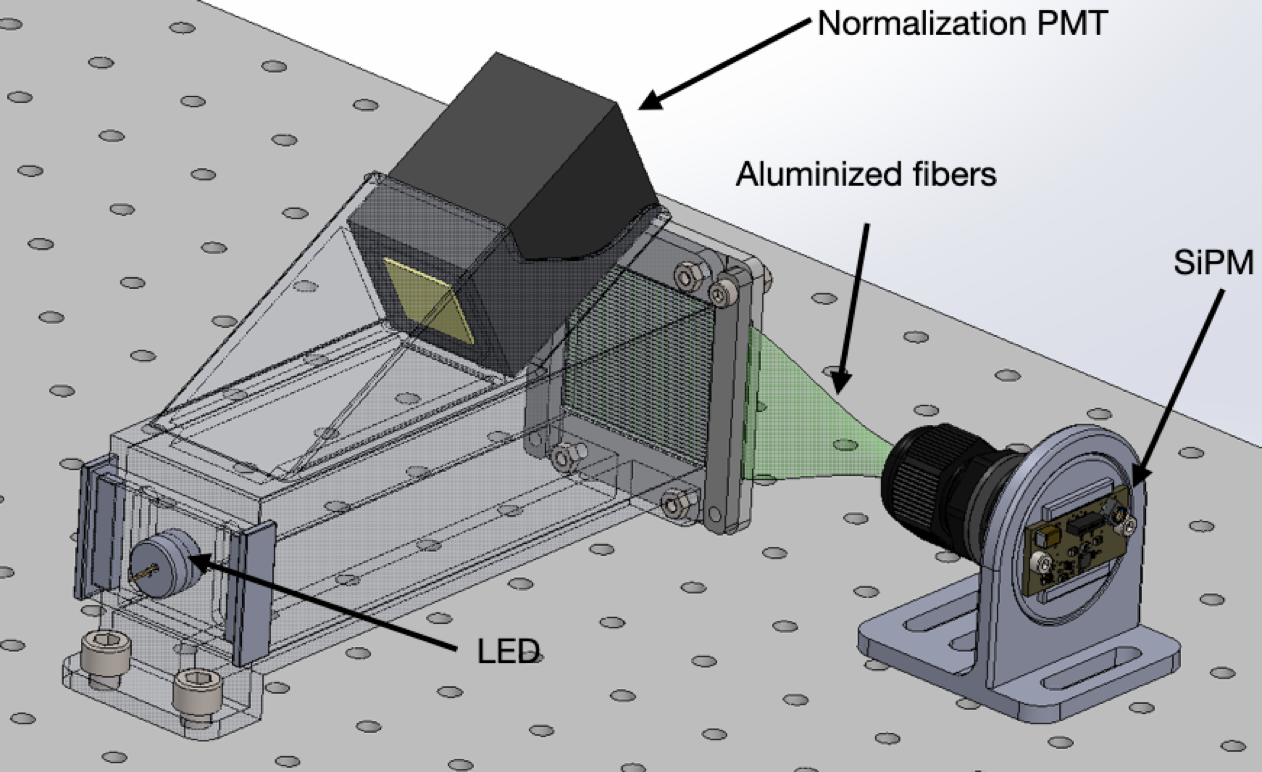}
\qquad
\includegraphics[width=.4\textwidth]{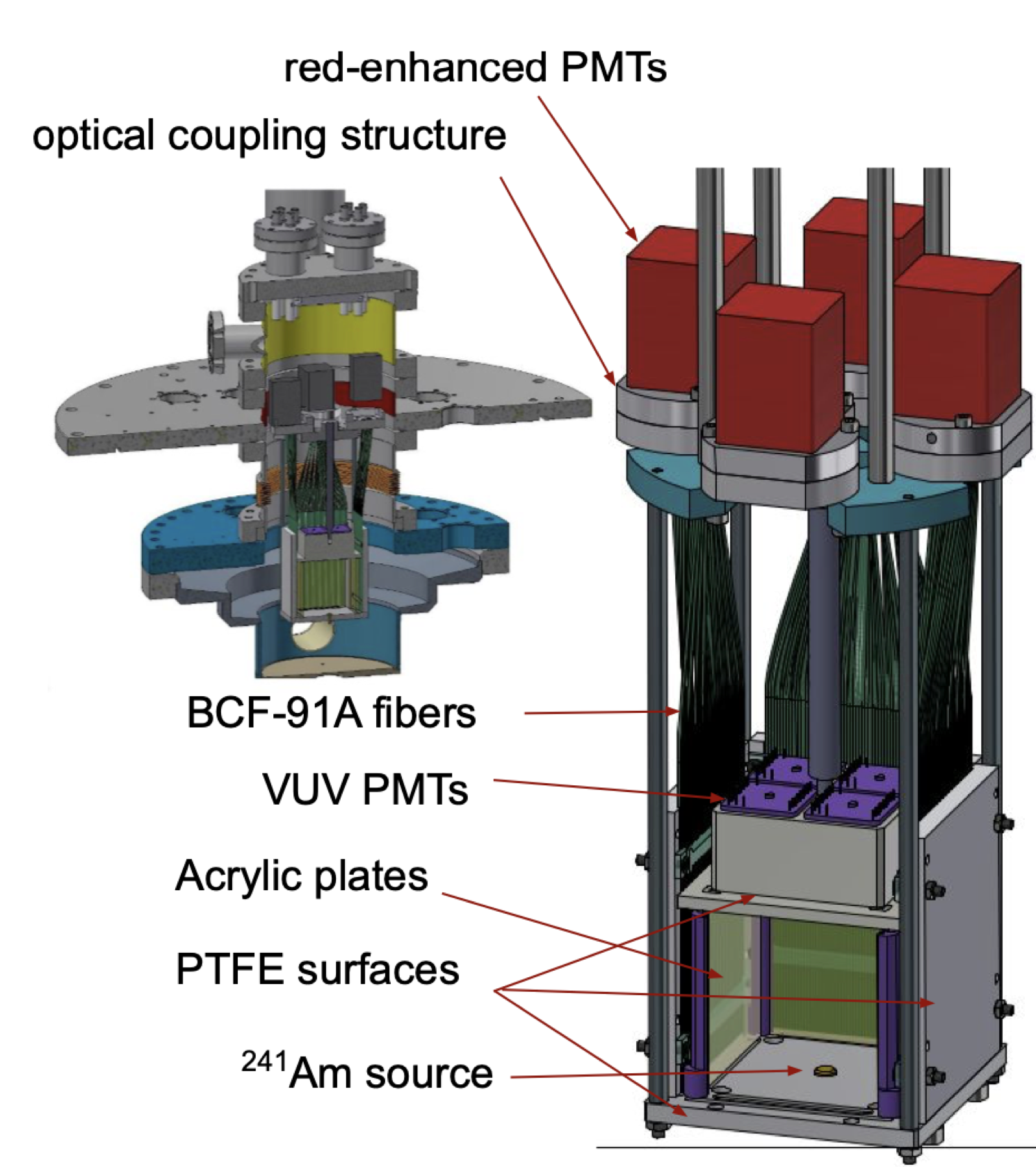}
\caption{\emph{Left}: table-top measurement of the light collection efficiency of WLS fibers read out by a SiPM. \emph{Right}: the FROGXe experiment at the Ben Gurion University of the Negev for the measurement of the light collection efficiency WLS fibers in xenon.\label{fig:dipc_setup}} 
\end{figure}

Several small-scale measurements have been conducted in order to measure the light collection efficiency of WLS fibers. In one experiment, whose setup is shown in figure~\ref{fig:dipc_setup} (left), a small bundle of 32 Y11 (B2) fibers, with a diameter of 1~mm, 7~cm long and aluminized on one end, is placed on a PTFE panel and illuminated with a 430~nm \cite{thorlabs_430}  (375~nm \cite{thorlabs_375}) LED, placed at a distance of 6~cm from the fibers. The light collected by the fiber is read out by a Hamamatsu S13360-6075CS SiPM \cite{hamamatsu}. The light collected is compared with the one detected by placing the photosensor directly in front of the LED at the same distance, normalizing by the area illuminated and by the photon detection efficiency of the SiPM. The amount of light emitted by the LED is monitored by a PMT, placed at a small angle above the LED. The light collection efficiency of the fibers coupled to the SiPM is approximately 3\%, in good agreement with a Geant4~\cite{GEANT4:2002zbu} simulation of the setup.

In the FROGXe experiment conducted at the Ben Gurion University of the Negev (figure \ref{fig:dipc_setup} right), a PTFE box is covered on four sides by 64 Luxium BCF-91A fibers~\cite{luxium} with a diameter of 1~mm. The fibers, which have absorption and emission spectra analogous to the Kuraray Y11 model, are aluminized on one end and read out by red-enhanced PMTs (H13543-20 \cite{hamamatsu_pmt}) on the other end. TPB can be evaporated on acrylic plates sitting in front of the fibers or directly on the fibers. The setup is designed to fit in a xenon chamber. In initial measurements performed in air and with uncoated fibers, light is emitted by a plastic scintillator placed on top of an $^{241}$Am radioactive source. The light collection efficiency in this case was measured to be approximately 1\%, in accordance with estimations based on simulations. This is lower than the 3\% obtained with the table-top measurement described above, which is expected given the quantum efficiency of the PMTs compared with the Hamamatsu SiPM ($\sim$20\% vs. $\sim$50\%). The final estimate will come from operation in xenon.

\section{Summary and outlook\label{sec:summary}}
A barrel of optical fibers coated with a wavelength-shifting material represents a realistic option for the measurement of the energy of the particles interacting in a high-pressure gaseous xenon TPC. Both the simulation and the preliminary table-top measurements show that the detector has the light collection efficiency required for the detection of $^{83\mathrm{m}}$Kr calibration events in a tonne-scale experiment. However, significant engineering challenges remain, namely the cooling of SiPMs in a high-pressure environment. The technology choice must be finalized after performing the same measurements with coated fibers in a xenon chamber and will be then validated in a small-scale prototype of NEXT-HD, HD-DEMO.

\acknowledgments

The NEXT Collaboration acknowledges support from the following agencies and institutions: the European Research Council (ERC) under Grant Agreement No. 951281-BOLD; the European Union’s Framework Programme for Research and Innovation Horizon 2020 (2014–2020) under Grant Agreement No. 957202-HIDDEN; the MCIN/AEI of Spain and ERDF A way of making Europe under grants PID2021-125475NB and the Severo Ochoa Program grant CEX2018-000867-S; the Generalitat Valenciana of Spain under grants PROMETEO/2021/087 and CIDEGENT/2019/049; the Department of Education of the Basque Government of Spain under the predoctoral training program non-doctoral research personnel; the Spanish la Caixa Foundation (ID 100010434) under fellowship code LCF/BQ/PI22/11910019; the Portuguese FCT under project UID/FIS/04559/2020 to fund the activities of LIBPhys-UC; the Israel Science Foundation (ISF) under grant 1223/21; the Pazy Foundation (Israel) under grants 310/22, 315/19 and 465; the US Department of Energy under contracts number DE-AC02-06CH11357 (Argonne National Laboratory), DE-AC02-07CH11359 (Fermi National Accelerator Laboratory), DE-FG02-13ER42020 (Texas A\&M), DE-SC0019054 (Texas Arlington) and DE-SC0019223 (Texas Arlington); the US National Science Foundation under award number NSF CHE 2004111; the Robert A Welch Foundation under award number Y-2031-20200401. Finally, we are grateful to the Laboratorio Subterr\'aneo de Canfranc for hosting and supporting the NEXT experiment.


\bibliographystyle{JHEP}
\bibliography{biblio.bib}

\end{document}